\documentclass[epj]{svjour}
\usepackage{graphicx}
\usepackage{amsmath}
\usepackage{siunitx}
\usepackage{caption}
\usepackage{subcaption}
\begin{document}
\newcommand{\Npg}{$^{14}$N(p,$\gamma$)$^{15}$O}
\title{Towards a comprehensive study of the $^{14}$N(p,$\gamma$)$^{15}$O astrophysical key reaction: Description of the experimental technique including novel target preparation}

\author{A. Compagnucci\inst{1,2}
\and
 A. Formicola\inst{3}
\and
 M. Campostrini\inst{4}
\and
 J. Cruz\inst{5}
\and 
 M. Aliotta\inst{6}
\and
 C. Ananna\inst{7,8}
\and 
 L. Barbieri\inst{6}
\and
 F. Barile\inst{9,10}
\and
 D. Bemmerer\inst{11}
\and
 A. Best\inst{7,8}
\and
 A.~Boeltzig\inst{11}
\and
 C. Broggini\inst{12}
\and
 C.G. Bruno\inst{6}
\and
 A. Caciolli\inst{12,13}
\and
 F. Casaburo\inst{14,15}
\and
 F. Cavanna\inst{16}
\and
 G.F. Ciani\inst{9}
\and
 P. Colombetti\inst{16,17}
\and
 P. Corvisiero\inst{14,15}
\and
 L. Csedreki\inst{18}
\and
 T. Davinson\inst{6}
\and
 R. Depalo\inst{19}
\and
 A. Di Leva\inst{7,8}
\and
 Z. Elekes\inst{18}
\and
 F. Ferraro\inst{1}
\and
 Zs. F\"ul\"op\inst{18}
\and
 A. Guglielmetti\inst{19}
\and
 C. Gustavino\inst{3}
\and
Gy. Gy\"urky\inst{18}\mail{Gy. Gy\"urky, gyurky@atomki.hu, A. Formicola, Alba.Formicola@roma1.infn.it}
\and
 G. Imbriani\inst{7,8}
\and
 M. Junker\inst{2}
\and
 M. Lugaro\inst{20,21}
\and
 P. Marigo\inst{12,13}\thanks{Deceased}
\and
J. Marsh\inst{6}
\and
 E. Masha\inst{11,19}
\and
 R. Menegazzo\inst{12}
\and
 V. Paticchio\inst{9}
\and
 D. Piatti\inst{12,13}
\and
 P. Prati\inst{14,15}
\and
 D. Rapagnani\inst{7,8}
\and
 V. Rigato\inst{4}
\and
 D. Robb\inst{6}
\and
 L. Schiavulli\inst{9,10}
\and
 R. S. Sidhu\inst{6,22}
\and
 J. Skowronski\inst{12,13}
\and
 O. Straniero\inst{23}
\and
 T. Sz\"ucs\inst{18}
\and
 S. Turkat\inst{12,13}
\and
 S. Zavatarelli\inst{14}
}                     
%
%
\institute{
Gran Sasso Science Institute, Viale F. Crispi 7, 67100, L'Aquila, Italy 
\and
Istituto Nazionale di Fisica Nucleare Laboratori Nazionali del Gran Sasso (LNGS), Via G. Acitelli 22, 67100 Assergi, Italy 
\and
Istituto Nazionale di Fisica Nucleare, Sezione di Roma, Piazzale A. Moro 2, 00185 Roma, Italy 
\and
Laboratori Nazionali di Legnaro, Viale dell’Universit\`a 2, I-35020, Legnaro (PD), Italy 
\and
LIBPhys, LA-REAL, Faculdade de Ciências e Tecnologia, NOVA FCT, Universidade NOVA de
Lisboa, Lisboa, Portugal 
\and
SUPA, School of Physics and Astronomy, University of Edinburgh, Peter Guthrie Tait Road, EH9 3FD Edinburgh, United Kingdom 
\and
Universit\`a degli Studi di Napoli ``Federico II'', Dipartimento di Fisica ``E. Pancini'', Via Cintia 21, 80126 Napoli, Italy 
\and
Istituto Nazionale di Fisica Nucleare, Sezione di Napoli, Via Cintia 21, 80126 Napoli, Italy 
\and
Istituto Nazionale di Fisica Nucleare, Sezione di Bari, Via E. Orabona 4, 70125 Bari, Italy 
\and
Universit\`a degli Studi di Bari, Dipartimento Interateneo di Fisica, Via G. Amendola 173, 70126 Bari, Italy 
\and
Helmholtz-Zentrum Dresden-Rossendorf, Bautzner Landstra\ss{}e 400, 01328 Dresden, Germany 
\and
Istituto Nazionale di Fisica Nucleare, Sezione di Padova, Via F. Marzolo 8, 35131 Padova, Italy 
\and
Universit\`a degli Studi di Padova, Via F. Marzolo 8, 35131 Padova, Italy 
\and
Istituto Nazionale di Fisica Nucleare, Sezione di Genova, Via Dodecaneso 33, 16146 Genova, Italy 
\and
Universit\`a degli Studi di Genova, Via Dodecaneso 33, 16146 Genova, Italy 
\and
Istituto Nazionale di Fisica Nucleare, Sezione di Torino, Via P. Giuria 1, 10125 Torino, Italy 
\and
Universit\`a degli Studi di Torino, Via P. Giuria 1, 10125 Torino, Italy 
\and
HUN-REN Institute for Nuclear Research (ATOMKI), PO Box 51, 4001 Debrecen, Hungary  
\and
Universit\`a degli Studi di Milano \& Istituto Nazionale di Fisica Nucleare, Sezione di Milano, Via G. Celoria 16, 20133 Milano, Italy 
\and
Konkoly Observatory, HUN-REN Research Centre for Astronomy and Earth Sciences, Konkoly Thege Miklós út 15-17, H-1121 Budapest, Hungary 
\and
ELTE Eötvös Loránd University, Institute of Physics, Pázmány Péter sétány 1/A, Budapest 1117, Hungary 
\and
School of Mathematics and Physics, University of Surrey, Guildford, GU2 7XH, United Kingdom 
\and
INAF Osservatorio Astronomico d'Abruzzo, Via Mentore Maggini, 64100 Teramo, Italy 
}
\date{Received: date / Revised version: date}
%
\abstract{
While the \Npg\ reaction plays a key role in the hydrogen-burning processes in various stellar conditions, its reaction rate is not known with sufficient precision. Therefore, the first scientific project at the recently launched Bellotti Ion Beam Facility of the Laboratori Nazionali del Gran Sasso was the measurement of the \Npg\ reaction cross section in the proton energy range between 250 and 1500\,keV.  
In this paper, the experimental techniques are summarized with special emphasis on the description of solid state nitrogen target production and characterization. The first results of the reaction yield measured at 55$^\circ$ detection angle are also presented.}
\maketitle

\section{Introduction}
\label{sec:intro}

The \Npg\ reaction is one of the key reactions in nuclear astrophysics playing a pivotal role in  stellar processes because it is the slowest and thus the most significant reaction of the CNO cycle of hydrogen burning \cite{Wiescher2010}. The CNO cycle has relevance in various fields of nuclear astrophysics. In the comprehensive description of our Sun (the standard solar model \cite{Salmon2021}), the rate of the \Npg\ reaction represents one of the largest uncertainties, making the model predictions less precise than some of the astronomical observations to be compared to. This is crucial in the present era, when solar neutrinos from the CNO cycle have been observed directly \cite{Borexino2020}. This reaction is also closely related to the determination of the age of the universe. One way of putting a lower limit to this is the age determination of old stellar globular clusters that formed soon after the Big Bang \cite{DEGLINNOCENTI200413}. The age of the globular clusters can only be determined accurately if the rate of the reaction \Npg\ is well known \cite{Imbriani2004}. 

From the point of view of nuclear physics, \Npg\ is a complex reaction: The capture of the proton in $^{14}$N proceeds through both direct capture and various resonances and involves several transitions in the $^{15}$O nucleus. As it is often the case in nuclear astrophysics, the cross section of \Npg\ is too low to be measured in the energy range relevant to stellar processes, making theory-based extrapolation to low energies inevitable. Such an extrapolation - often carried out with the R-matrix approach \cite{Descouvemont2010} - necessitates the precise knowledge of the differential reaction cross sections in a wide energy range for all the possible transitions. Any uncertainty in the higher energy cross sections, or the missing information for some partial cross sections, will directly influence the accuracy of the low energy extrapolations and thus the calculation of the astrophysical reaction rate \cite{Frentz2022}. 

Since the dawn of nuclear astrophysics, the importance of the \Npg\ reaction has triggered many experimental investigations. The first detailed and comprehensive study was carried out by Schr\"oder \textit{et al.} \cite{SCHRODER1987240}, who measured the partial cross sections for all the observable transitions and provided cross section extrapolations down to astrophysical energies (a few tens of keV, depending on the astrophysical site). Owing to the outdated experimental techniques and some incorrectly considered systematic uncertainties, the results of \cite{SCHRODER1987240} have a relatively high uncertainty and in some cases the data had to be ex post corrected for the true coincidence summing effect \cite{Imbriani2005}. In the last two decades, many experiments were dedicated to this reaction (for a complete list and references see the recently published review \cite{acharya2024solarfusioniiinew}). These new experiments mostly concentrated on the lowest measurable energies and on the transitions having the highest astrophysical relevance. A comprehensive experimental data set in a wide energy range, including all possible transitions and also providing angular distribution information is, however, still missing since the work of Schr\"oder \textit{et al.} \cite{SCHRODER1987240}. 

Due to the insufficient experimental information and resulting ambiguities of the low-energy extrapolations, the recently published Solar Fusion III review \cite{acharya2024solarfusioniiinew} assigns an 8.4\,\% uncertainty to the zero energy extrapolated total S-factor\footnote{The astrophysical S-factor is related to the reaction cross section but removes its strong energy dependence due to the Coulomb-barrier penetration. For its definition see \cite{Iliadis2015}.}. Owing to a more careful analysis of the available experimental data, this value is higher than the 7\,\% uncertainty published in the previous edition of the same review \cite{Adelberger2011}, and clearly worse than the precision of some related astronomical observations, like e.g. some of the solar neutrino fluxes or light element abundances \cite{Xu2023,Christensen-Dalsgaard2021}. Therefore,  a high priority recommendation in ref. \cite{acharya2024solarfusioniiinew} is a new study of the \Npg\ reaction.

Answering to this need, the first scientific program of the recently launched Bellotti Ion Beam Facility (BIBF) of the Laboratori Nazionali del Gran Sasso is the measurement of the \Npg\ cross section. Exploiting the conditions provided by the deep underground facility, the differential cross section for primary transitions to the \num{6.79}~MeV state and ground state, as well as the secondary transitions from the \num{6.79}, \num{6.18}, \num{5.24} and \num{5.18}~MeV states were measured in the energy range between \num{0.25} and \num{1.5} MeV at five detection angles. 

In this paper we focus on the details of the experimental technique and we introduce some first results to show the capability and sensitivity of the setup. Target preparation and characterization techniques are discussed in detail in Sects. \ref{sec:target} and \ref{sec:target_char}, respectively. Information about the accelerator, the target chamber and the detection setup is given in Sect. \ref{sec:experimental}. In Sect. \ref{sec:results} reaction yield results obtained at 55$^\circ$ detection angle are presented as an example. The conclusions and outlook are given in Sect. \ref{sec:conclusions}.

\section{Target preparation}
\label{sec:target}
The \Npg\ reaction cross-section measurements  required targets that were optimized for measuring weak signals from the resonant and direct capture components of this reaction at energies below 1500\,keV. To reduce target-related systematic uncertainties, two different methods of target preparation were used: ion-implantation and sputtering. 

Both techniques are widely used in thin films and surface modification applications with advantages and disadvantages. Ion implantation offers the possibility to produce isotopically pure $^{14}$N targets, while the reactive sputtering technique uses a nitrogen gas isotopically enriched in $^{14}$N to 99.99\%, which inevitably will leave some residual traces of $^{15}$N in the nitride layer. On the other side, ion implantation requires a predictive design of the implantation profile (described in Sect. \ref{sec:implanted}), while in the case of reactive magnetron sputtering the composition can be controlled very precisely with state of the art plasma optical emission methods, and once calibrated, it is highly reproducible and allows for the high throughput production of layers of different thickness (as described in Sect. \ref{sec:sputtered}).

Tantalum was chosen as backing material since (i) it is possible to form a stable nitrogen compound by nitride formation or ion implantation at room temperature and, (ii) for the energy range of interest, Ta does not produce $\gamma$-radiation upon proton bombardment that could interfere with our measurements. It is commercially available in a high purity state, thus avoiding contaminants that could act as a source of beam-induced background (BIB) radiation, and it obeys the stringent LNGS neutron production limits.

The sputtered TaN layer thicknesses and the $^{14}$N implantation energy of 40 keV were chosen to yield targets with energetic thickness of 20-40\,keV at a proton beam energy of 280\,keV, thus satisfying both effective beam energy and counting rate requirements for this experiment. 

\subsection{Ta backing preparation}
\label{sec:backings}

Ta foils of 0.25-mm thickness and commercial purity $\geq 99.9\%$ were laser cut into 40.5-mm-diameter disc backings, and then etched using an acid solution (nitric acid and hydrochloric acid 10:1, electronic grade, from CARLO ERBA Reagents) at a temperature of \num{80}$^\circ$C to remove surface contaminants that could act as a source of BIB signals. The etching process residues were removed by rinsing the Ta discs with distilled water.

\subsection{Target Implantation with the Danfysik S1090 ion implanter}
\label{sec:implanted}

Monte Carlo (MC) simulations were performed using the SRIM2013 code \cite{ZIEGLER20101818} to select the most suitable energy and fluence for $^{14}$N implantations, aiming to achieve a target with the desired properties. Predicted implantation profiles were determined based on the implanted-versus-sputtered balance equation~\cite{ionimp}:


\begin{equation}
\begin{split}
 & N(x)  = N  \frac{1- \textup{BS}}{2S} \times  \\  & \left( \textup{erf} \left( \frac{x - R_p +  D_n(S/N)}{\sqrt{2}\Delta R_p}\right) -\textup{erf}\left(\frac{x - R_p }{\sqrt{2}\Delta R_p} \right) \right) ,
 \end{split}
\label{eq:impl}
\end{equation}
where $N(x)$ is the $^{14}$N atomic density for depth $x$, $N$ the Ta atomic density (\num{5.525e22}~atoms/cm$^3$), $D_n$ the $^{14}$N nominal fluence in atoms/cm$^2$, $R_p$ (\num{41.4}~nm) and $\Delta R_p$ (\num{22.2}~nm) the range and straggling of $^{14}$N ions in tantalum, $S$ (\num{0.91}~atoms/ion) the sputtering yield, and $BS$ (\num{0.2114}) the fraction of $^{14}$N ions that are not implanted but backscattered out of the tantalum.  
The values of $R_p$, $\Delta R_p$, $S$ and $BS$ were obtained directly from the MC SRIM simulation. Eq.~\ref{eq:impl} is based on the assumptions that the sputtering yield remains constant and is identical for both substrate and implanted ions, that knock-on effects are negligible, and that radiation-induced volume changes can be ignored. Additionally, it does not account for saturation or diffusion effects, which may limit its accuracy in predicting implantation depth profiles at very high fluences. Nevertheless, it serves as a useful starting point.

\begin{figure}[htb]
	\centering 
	\includegraphics[width=0.5\textwidth]{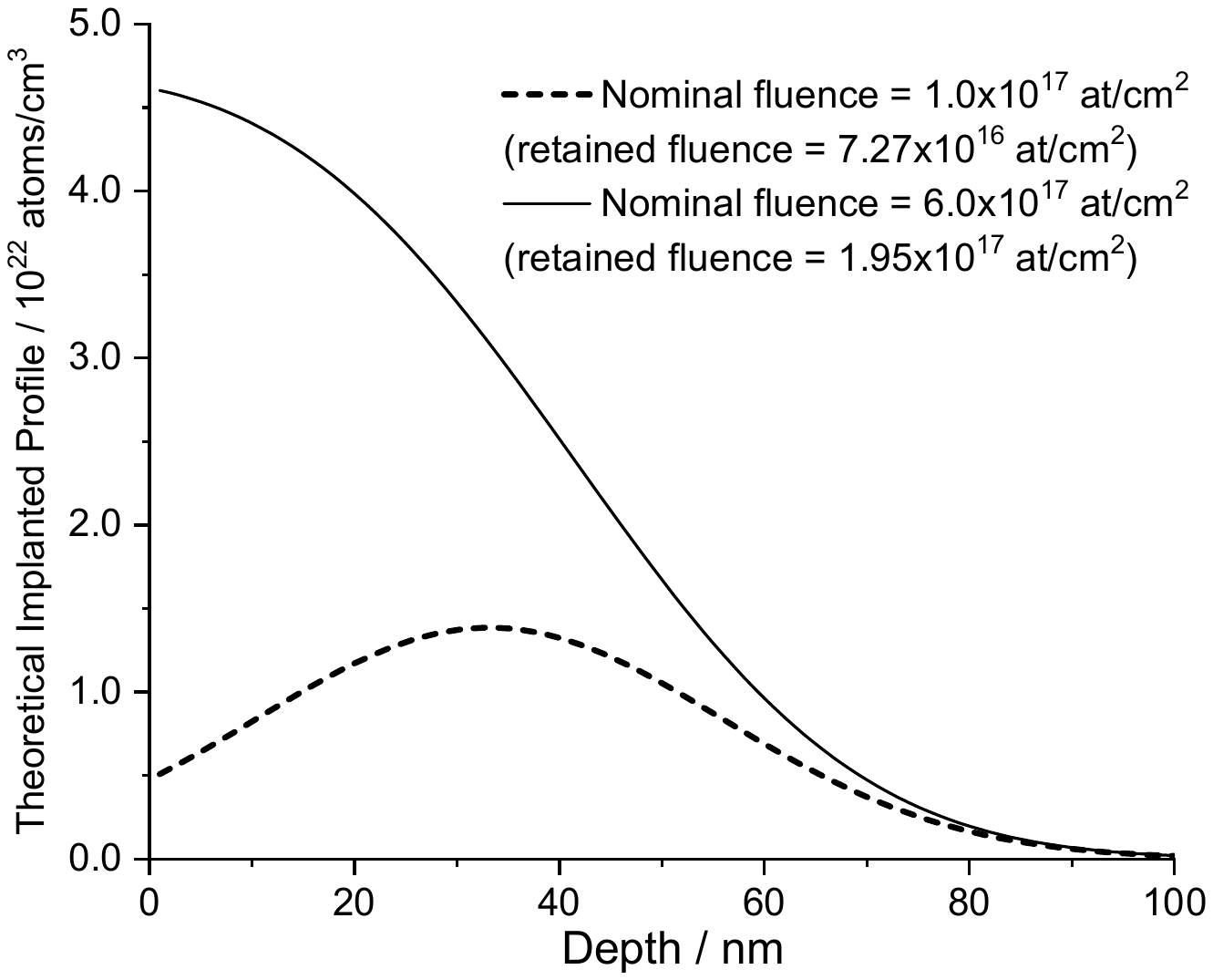}	
	\caption{Predicted $^{14}$N depth profiles for two different nominal fluences. The $^{14}$N implantation energy is 40\,keV.} 
	\label{fig:implantation_profile}%
\end{figure}

Figure~\ref{fig:implantation_profile} shows the curves given by Eq.~\ref{eq:impl} for two different nominal fluences of $^{14}$N ions implanted with an energy of \num{40}~keV. 
For each nominal fluence, the retained fluence is computed by integrating the corresponding atomic density curve along depth. Figure~\ref{fig:implantation_profile} shows that for an energy of \num{40}~keV, the $^{14}$N depth profile saturates for a nominal fluence of \num{6.0e17}~atoms/cm$^2$, with a corresponding retained fluence of \num{1.95e17}~atoms/cm$^2$.

\begin{figure}[htb]
	\centering 
	\includegraphics[angle=-90,width=0.49\textwidth]{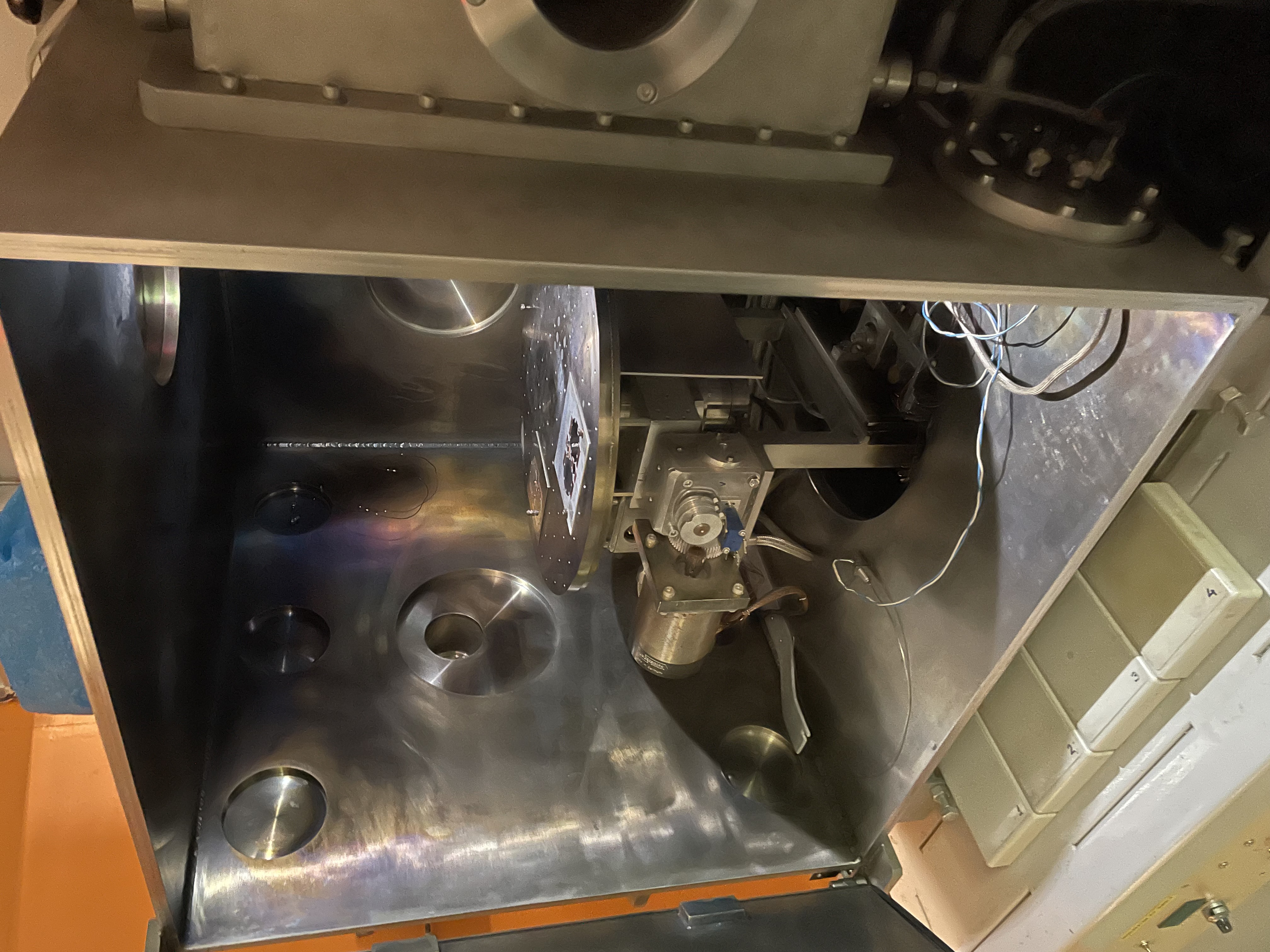}
 \caption{Tantalum (Ta) discs mounted on the rotating holder inside the vacuum chamber of the ion implanter, prior to implantation with  $^{14}$N ions.}
	\label{fig:implanter_photo}%
\end{figure}

Before implantation, the Ta backings were hand polished with SiC grinding paper (grid \num{2500} and \num{4000}) and cleaned with ethanol.
\num{40}~keV $^{14}$N$^{++}$ ions were implanted using the high flux \num{210}~kV ion implanter, model S1090 from Danfysik operated at CTN-IST (Portugal)~\cite{alves_insider_2021}, see Fig.~\ref{fig:implanter_photo}. 
This system provides a mass resolution of $M/\Delta M > 150$ to 250 (where $\Delta M$ is the FWHM of the mass peak) and is capable of operating at energies of 1-200~keV with maximum target ion currents of $\sim$500~$\mu$A for $^{14}$N$^{++}$ and $\sim$100~$\mu$A for $^{14}$N$^+$. During implantation, the pressure was kept at $\approx 10^{-6}$~mbar. 

Two sets of implantations were performed at room temperature: one with a nominal fluence of \num{6.0e17}~atoms/cm$^2$ (current density = \num{8.1}\,$\mu$A/cm$^2$) and another with twice that fluence (current density = \num{6.9} $\mu$A/cm$^2$). The first value corresponds to the predicted saturation profile (Fig.~\ref{fig:implantation_profile}), while the second was used to investigate whether further $^{14}$N density increase was possible (see also Sect.~\ref{sec:target_char_impl}).


\subsection{TaN sputtered targets}
\label{sec:sputtered}

\subsubsection{Sputtering system description}

\begin{figure*}[tb]
	\centering
    \begin{subfigure}[b]{0.49\textwidth}
         \centering
         \includegraphics[width=0.9\textwidth]{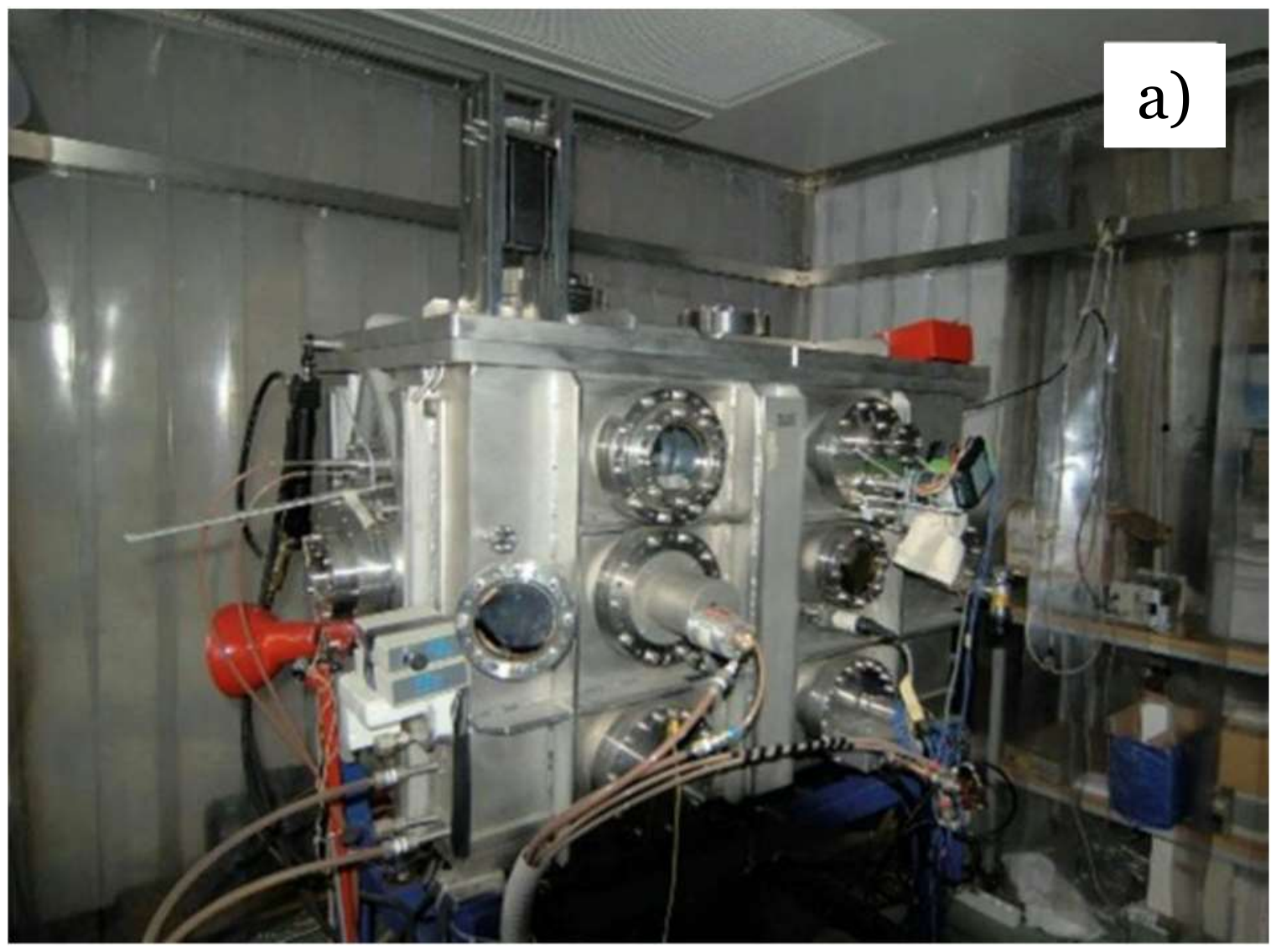}
         \caption{PVD magnetron sputtering system}
         \label{fig:magnetron}
     \end{subfigure}
\hfill
     \begin{subfigure}[b]{0.49\textwidth}
         \centering
         \includegraphics[width=0.9\textwidth]{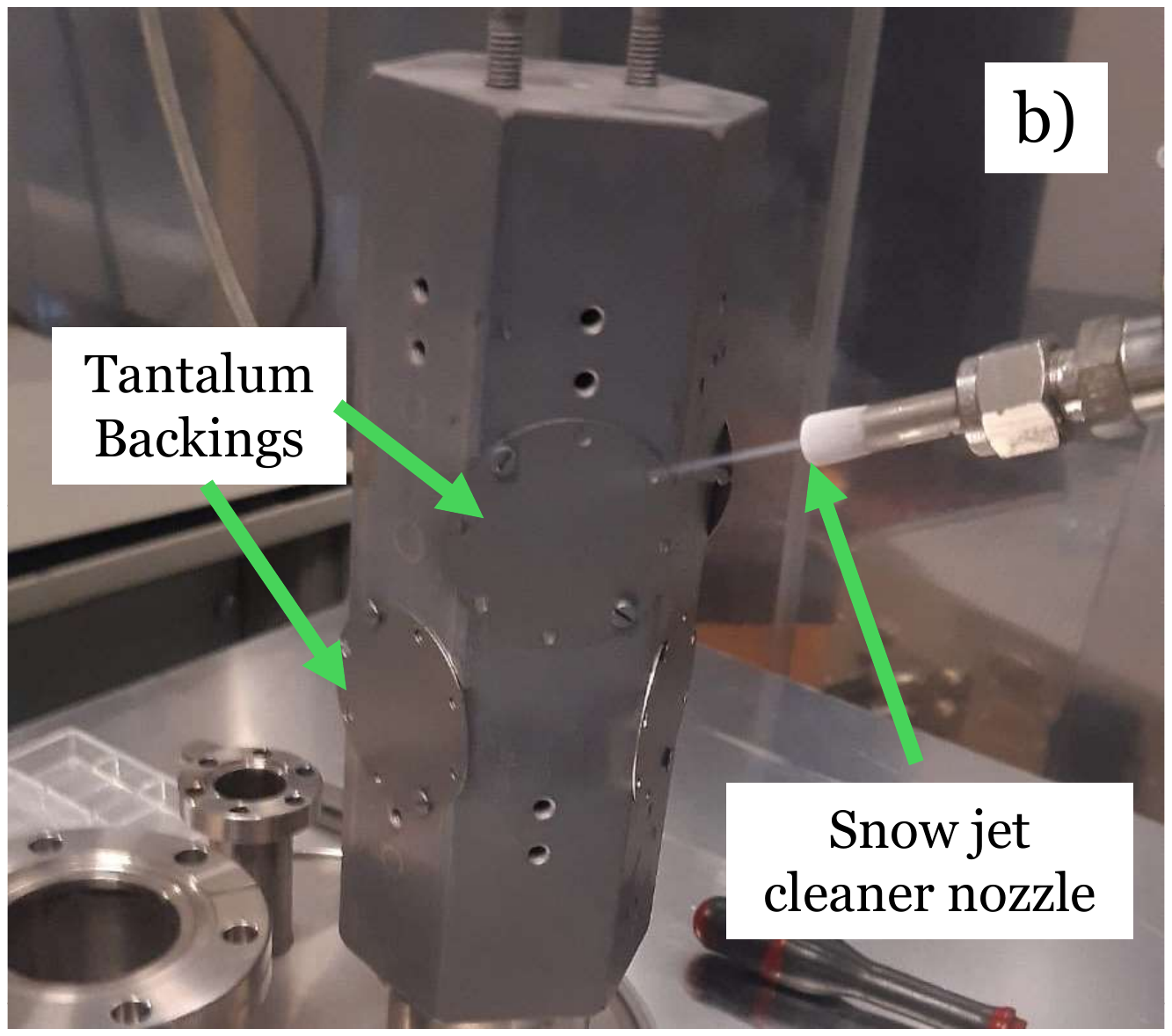}
         \caption{Hexagonal sample holder during CO$_2$ cleaning procedure}
         \label{fig:cleaning}
     \end{subfigure}
\hfill
	\caption{Magnetron sputtering system used to produce the TaN targets. b) a picture of the sample holder with Ta backing disks mounted and the snow jet cleaner nozzle used to clean their surface prior to deposition.  The equipment is positioned in a filtered air laminar flow enclosure to reduce dust uptake.} 
	\label{fig:sputtering_system}%
\end{figure*}

Tantalum nitride targets were produced by reactive magnetron sputtering at the INFN, Laboratori Nazionali di Legnaro. 
The sputtering apparatus was upgraded with various devices to ensure the high quality and precise composition required for the experiment.

It consists of a large vacuum chamber (Fig.~\ref{fig:magnetron}) equipped with two facing sputtering sources (AJA International mod. 2056) in an unbalanced closed magnetic field configuration for the target production, and a third sputtering source (the same type) for the active gettering system.

The sputtering material was tantalum with \num{99.99}\% purity. The sample holder (Fig.~\ref{fig:cleaning}) is a solid aluminum hexagonal prism positioned between the two sputtering sources, at a deposition distance of 10~cm. 
It remains in constant rotation and is negatively biased throughout the process. 
Given the chamber size, thermal baking at \num{120}$^\circ$C is not feasible. 
Instead, an active getter system was developed: the third magnetron sputtering source is positioned opposite to the sample holder and is used to sputter high purity (\num{99.95}\%) metallic zirconium onto a slowly rotating disk. 
This setup continuously provides fresh Zr metal onto a large area, promoting the gettering of residual gases other than Ar within the chamber. 
Active gettering is performed both before and during the deposition process to remove outgassing contaminants as internal surfaces heat up.

\begin{figure*}[tb]
	\centering
    \begin{subfigure}[b]{0.49\textwidth}
         \centering
         \includegraphics[width=\textwidth]{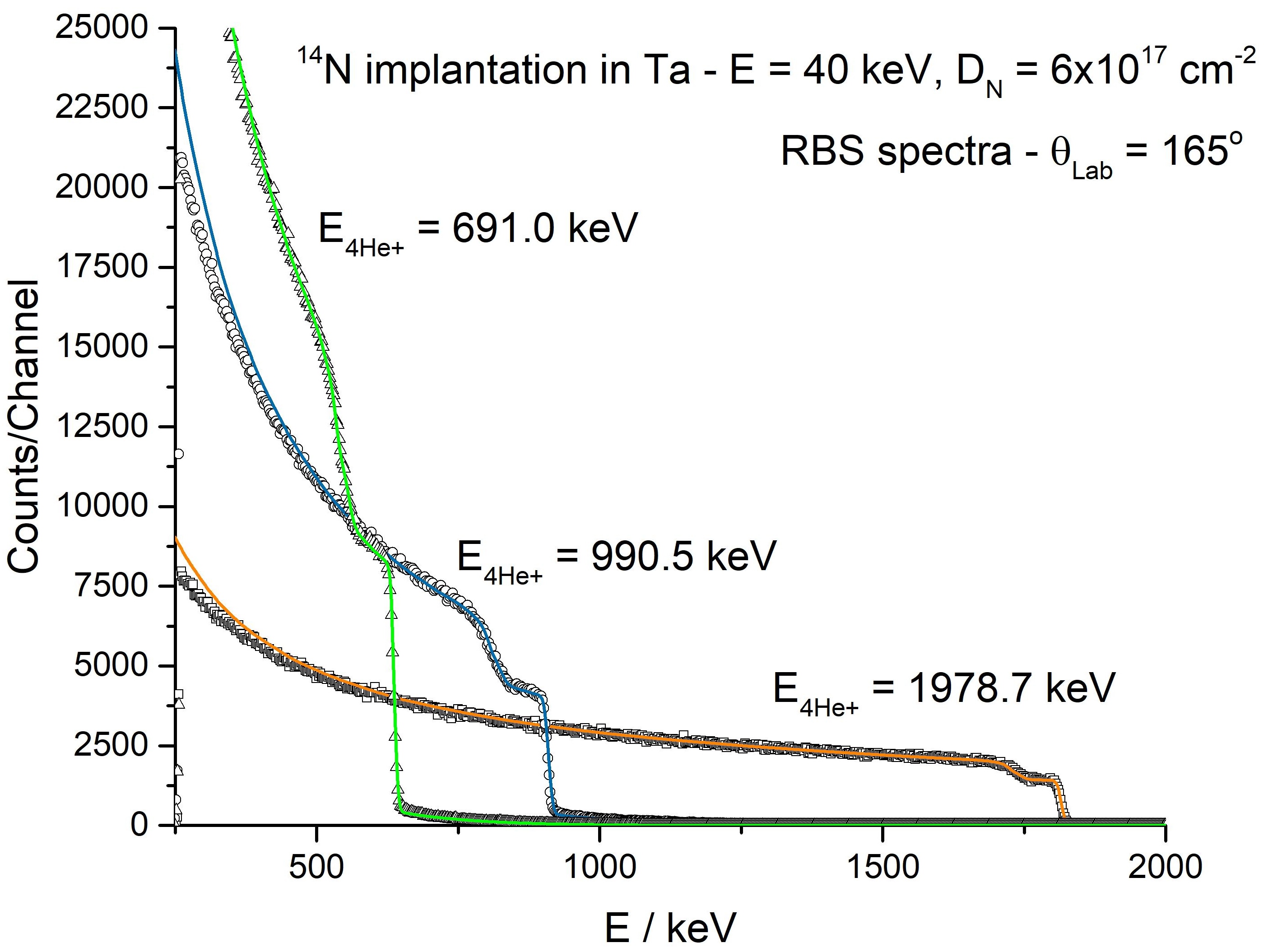}
         \caption{$^4$He$^+$ backscattering spectra}
     \end{subfigure}
\hfill
     \begin{subfigure}[b]{0.49\textwidth}
         \centering
         \includegraphics[width=\textwidth]{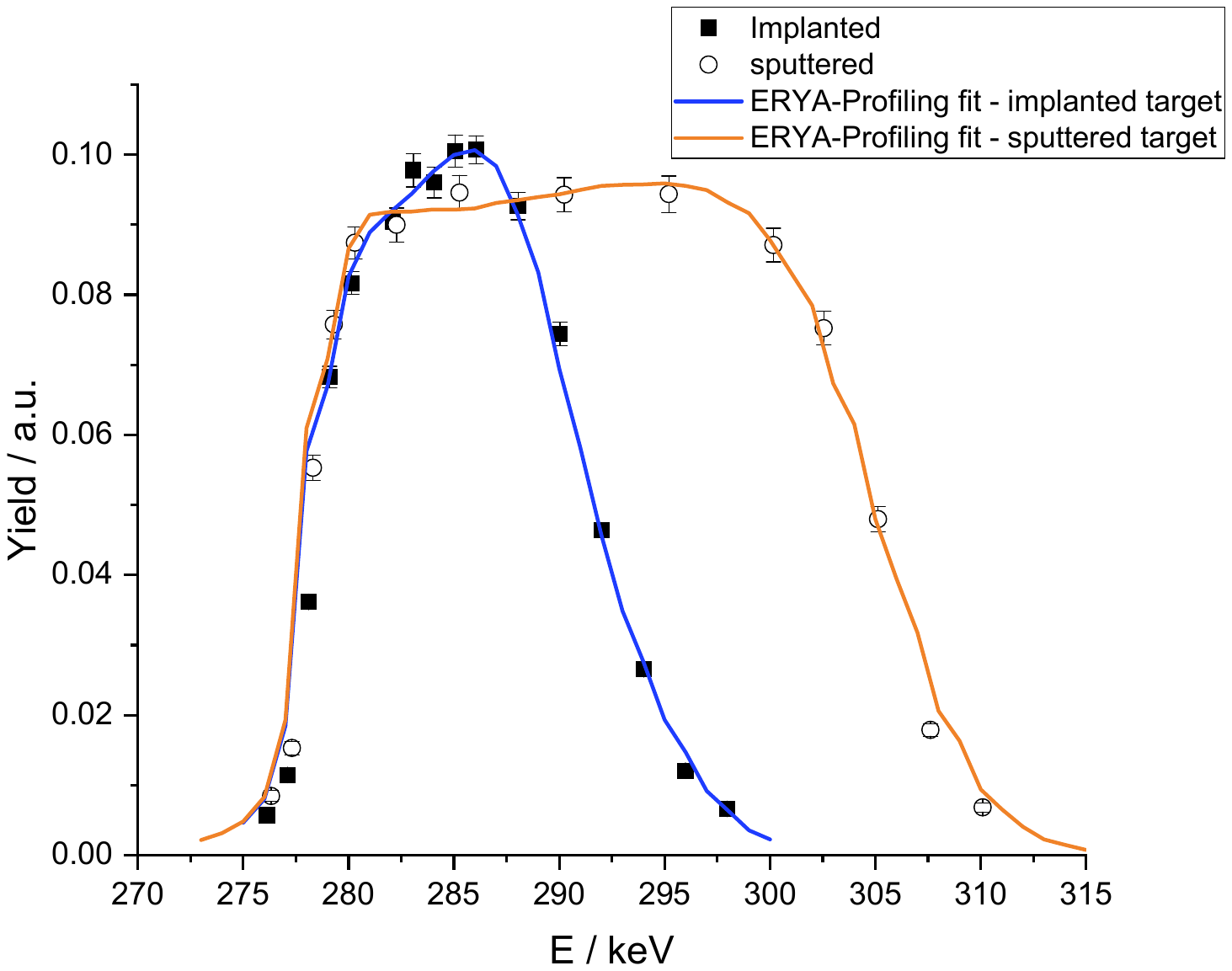}
         \caption{Scans of the 278\,keV resonance}
     \end{subfigure}
\hfill
	\caption{
 On the left, the $^4$He$^+$ backscattering spectra (black points) for the implanted target ($D_n = \num{6.0e17}$~atoms/cm$^2$) measured at $\theta_{\textup{lab}}=\ang{165}$, for three different energies. The NDF simulated spectra are represented with solid lines. On the right, resonance scans measured at the LUNA-400 accelerator using the \num{278} keV resonance in \Npg\ for test targets produced with sputtering and implantation, before irradiation at BIBF. ERYA-Profiling simulations \cite{MANTEIGAS2022108307} for each target are also plotted as solid lines.} 
	\label{fig:RBS_and_scan}%
\end{figure*}

An optical emission monitor system precisely regulates the reactive gas flow required for nitrides production. 
Using a piezo valve, it controls the flow of isotopically enriched  \num{99.99}\% $^{14}$N gas (Merck). 
Additionally, a mass-flow meter installed in series allows for gas-consumption monitoring. 
The entire gas injection line is composed of all-metal components and Swagelok VCR connections.

\subsubsection{Sputtering process description and parameters}

The sample holder accommodates up to five tantalum backing disks per deposition cycle, with an additional silicon substrate included as a witness sample for coating characterization. 
Sample mounting is performed in a laminar flow clean room (ISO6), and each sample is pre-treated (before installation into the sputtering chamber) with a CO$_2$ snow jet cleaner to remove potential environmental dust residues (Fig.~\ref{fig:cleaning}).
After sample installation, the system is pumped down to reach a base pressure below \num{2.0e-6}~mbar.

Prior to deposition, an Ar plasma cleaning treatment is performed on both the sputtering targets and samples. 
The chamber is filled with high-purity (N60) argon gas to reach the process pressure of \num{4.5e-3}~mbar, with the Ar gas flow managed by a mass flow controller (MKS 1179), and the pressure is measured using a capacitance manometer (MKS Baratron 627). 
Ar plasma cleaning removes the natural passivation oxide from both sputtering targets as well as from the Ta backing surfaces. 
During the target cleaning cycle the sample holder is connected to a pulsed-DC power supply operated at 200 kHz at a power of \num{80}~W (ENI - RPG-50). 
The sputtering sources are powered by an RF (\num{13.56}~MHz) power supply for tantalum (AE Dressler Cesar) and a pulsed-DC supply for zirconium (AE - Solvix Magix), each set at \num{350}~W.
After 30 minutes of Ar plasma cleaning, the sputtering chamber is re-evacuated. 
Thanks to the getter cycle the new base pressure typically drops below \num{1.0e-7}mbar. 
If this does not happen the cleaning procedure is repeated.

Once the cleaning phase is completed, the sample holder is biased using a DC power supply (KEPCO mod. BOP 100) at \num{-20}~V to promote low energy ion-assisted film growth. 
The chamber is then filled again with argon gas, to the pressure of \num{4.5e-3}~mbar.

At the end of the preparation phase, the sputtering system is ready for depositing the actual nuclear target material. 
The final coating consists of a bi-layer made of a high purity tantalum interlayer (\num{1.5e18} $\pm$ 5\,\% atoms/cm$^2$) covered by tantalum nitride as the target material for the cross-section measurements.
This dual-layer structure was chosen to reduce beam-induced background since the sputtered deposited Ta has less low-level impurities than the laminated one. 
The sputtered Ta interlayer also serves as an adhesion layer.

After the Ta interlayer deposition, the reactive gas injection system based on the Plasma Optical Emission closed-loop controller is activated to form the nitride layer. 
The optical emission system includes two main devices. 
The first is a monochromator (Verity Instruments Mod. EP200, band pass\,=\,4~nm), which measures the intensity of a single wavelength emission from neutral atoms and ions in the plasma, through a shielded high-vacuum collection optics and a quartz optical fiber, and converts it to a properly normalized voltage signal.  
In our case, the working wavelength was \num{243\pm2}~nm~\cite{NIST_ASD}, chosen on the basis of experimental tests to avoid interferences with the argon and nitrogen spectral lines. 
The second device is a PID closed loop controller, set to quickly adjust the piezo valve opening to maintain the desired line intensity (i.\,e. the tantalum erosion rate) when nitrogen is admitted in the sputtering chamber. 
The intensity of the Ta plasma emission line decreases with the admission of the reactive gas due to the reduction of the sputtering rate caused by nitrogen poisoning of the tantalum source~\cite{BERG2005215}.
Unlike other nitrides (e.g.\, TiN, ZrN), tantalum nitride sputtered films exist in crystallographic phases with different stoichiometries such as TaN, Ta$_2$N, Ta$_5$N$_6$~\cite{LU2001356}. To ensure the TaN coating stoichiometry, the emission line intensity was set to 45\% of that of the pure tantalum. This corresponded to a total working pressure of about \num{4.8e-3}~mbar (Ar + N$_2$) and a nitrogen flow rate of approximately 8~sccm.

\section{Target characterization}
\label{sec:target_char}

Accurate knowledge of the target stoichiometry, thickness, and stability under beam bombardment is crucial for absolute cross-section measurements. These properties were characterized using Ion Beam Analysis (IBA) techniques, specifically Rutherford Backscattering Spectrometry (RBS), Nuclear Reaction Analysis (NRA) with $(d,p)$ reactions, and Particle-Induced Gamma-ray Emission (PIGE). The analyses were conducted at four laboratories equipped with MV-scale particle  accelerators: ATOMKI (Debrecen, Hungary), CTN-IST (Lisbon, Portugal), LNL (Legnaro, Italy), and LNGS (Assergi, Italy).


\subsection{Implanted targets}
\label{sec:target_char_impl}

RBS spectra were obtained under vacuum conditions ($P = 5 \times 10^{-6}$ mbar) at the RBS beamline of the 2.5 MV Van de Graff accelerator at CTN-IST \cite{alves_insider_2021} using a $^4$He$^+$ beam and three different incident energies (691.0 keV, 990.5 keV, and 1978.7 keV). The RBS spectra were collected by a PIN photodiode from Hamamatsu with a 20 keV resolution in IBM geometry placed at 165$^\circ$ to the beam direction and covering a solid angle of 11.4 $\pm$ 0.8 msr. PIGE measurements were carried out at the LUNA-400 accelerator at the Gran Sasso National Laboratories (LNGS) \cite{formicola_luna_2003} with the profile measurement of the 278 keV resonance in $^{14}$N(p,$\gamma$)$^{15}$O. In both cases, targets were mounted perpendicular to the beam direction. RBS spectra were simultaneously fitted using the NDF code \cite{10.1063/1.119524}, including the beam straggling calculated by the Chu correction with the Tsch\"{a}lar effect, the double scattering contribution and the pulse pile-up effect (considering the Molodtsov and Gurbich algorithm). The deficiency method was applied, i.e., the nitrogen depth profile was extracted from the reduction of the backscattered yields from tantalum atoms induced by the change in stopping cross section owing to the nitrogen atoms. PIGE yields were simulated with the ERYA-Profiling code (Emitted Radiation Yield Analysis \cite{MANTEIGAS2022108307}) assuming a gaussian beam straggling. The $^{14}$N(p,$\gamma$)$^{15}$O excitation function at the resonance was parametrized with the function:
\begin{equation}
	\begin{split}
		& \sigma(E)  =  \frac{C \, (A + 1) \, \omega \gamma \, \Gamma}{A \, E \, \left[4 (E - E_R)^2 + \Gamma^2 \right]} +  \\  &  \frac{S_0 \exp(- 2 \, \pi \eta)}{E} ,
	\end{split}
	\label{eq:lorentz}
\end{equation}
where $A$ = 14, $\omega \gamma$ = 13.1 meV, $\Gamma$ = 1.12 keV, $E_R$ = 278 keV, $C$ = 2.607$\times 10^6$ keV mb (normalization factor to present the $\sigma(E)$ in units of mb), $S_0$ = 2.783$\times 10^6$ keV mb, and $\eta$ is the Sommerfeld parameter \cite{Adelberger2011}.

\begin{figure}[htb]
	\centering 
	\includegraphics[width=0.49\textwidth]{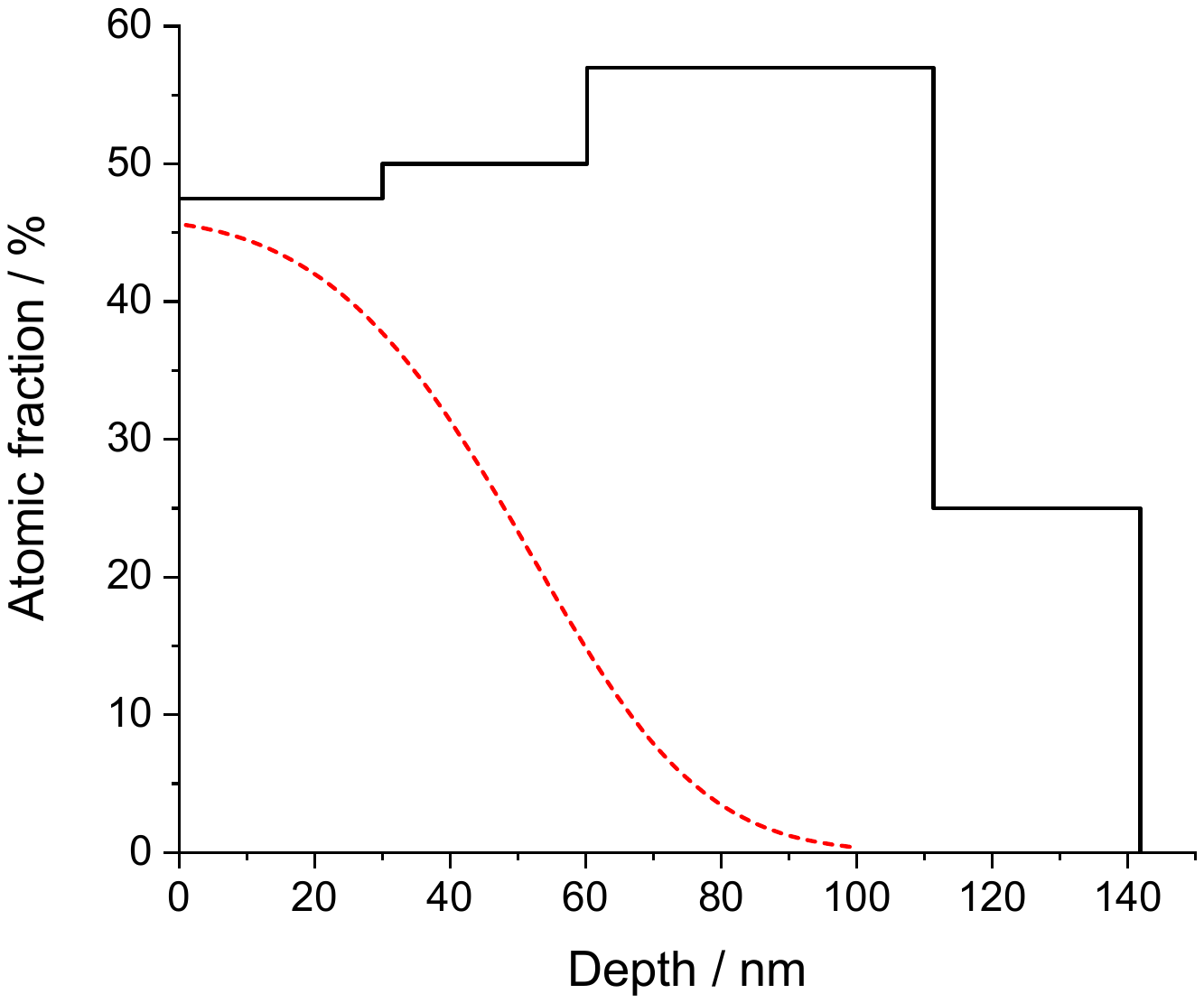}
 \caption{Nitrogen depth profiles and layer structure obtained by NDF+PIGE analysis and by Eq.~\ref{eq:impl}.}
	\label{fig:NDF-PIGE}%
\end{figure}

Figure~\ref{fig:RBS_and_scan}(a) presents the measured RBS spectra (black symbols) along with the corresponding NDF fit (lines) for the implantation with a nominal fluence of \num{6e17}~atoms/cm$^2$. Figure~\ref{fig:RBS_and_scan}(b) presents the 278 keV resonance scan performed at LUNA-400 accelerator (black points) plus the ERYA-Profiling simulation \cite{MANTEIGAS2022108307} (blue line). 
The combined RBS and PIGE analysis resulted in the $^{14}$N depth distribution presented in Fig.~\ref{fig:NDF-PIGE} (solid line), with the saturated profile of Fig.~\ref{fig:implantation_profile} superimposed (dashed line).
While the depth profile from Eq.~\ref{eq:impl} is smooth because it is based on a continuous function, the binning due to depth resolution limits in RBS+PIGE measurements means that atomic fraction values are averaged over discrete depth intervals, making the experimental profile appear more step-like than it truly is. Comparing both curves, we conclude that the depth profile obtained experimentally by PIGE+EBS qualitatively confirms the predicted implantation distribution of $^{14}$N in tantalum, though differences in peak position, width, and intensity suggest contributions from ion beam straggling, potential diffusion, or surface effects not captured by the theoretical model. Regarding the retained fluence, the experimental results yielded \num{3.3e17} atoms/cm$^2$, which is 1.7 times higher than the predicted value.

Similarly, RBS analysis results for the \num{12e17} atoms/cm$^2$ nominal fluence implantation showed a continuous drop of $^{14}$N concentration from the surface where the concentration was determined to be extremely high ($\sim$72\% at.), pointing to the possible formation of nitrogen gas bubbles. The calculated retained fluence amounted to a slight increase (3.5\%) with respect to the lower fluence implantation, thus showing that it is unnecessary to go beyond the nominal fluence of \num{6.0e17} atoms/cm$^2$.

\subsection{Sputtered targets}

\begin{figure}[htb]
	\centering 
	\includegraphics[width=0.45\textwidth]{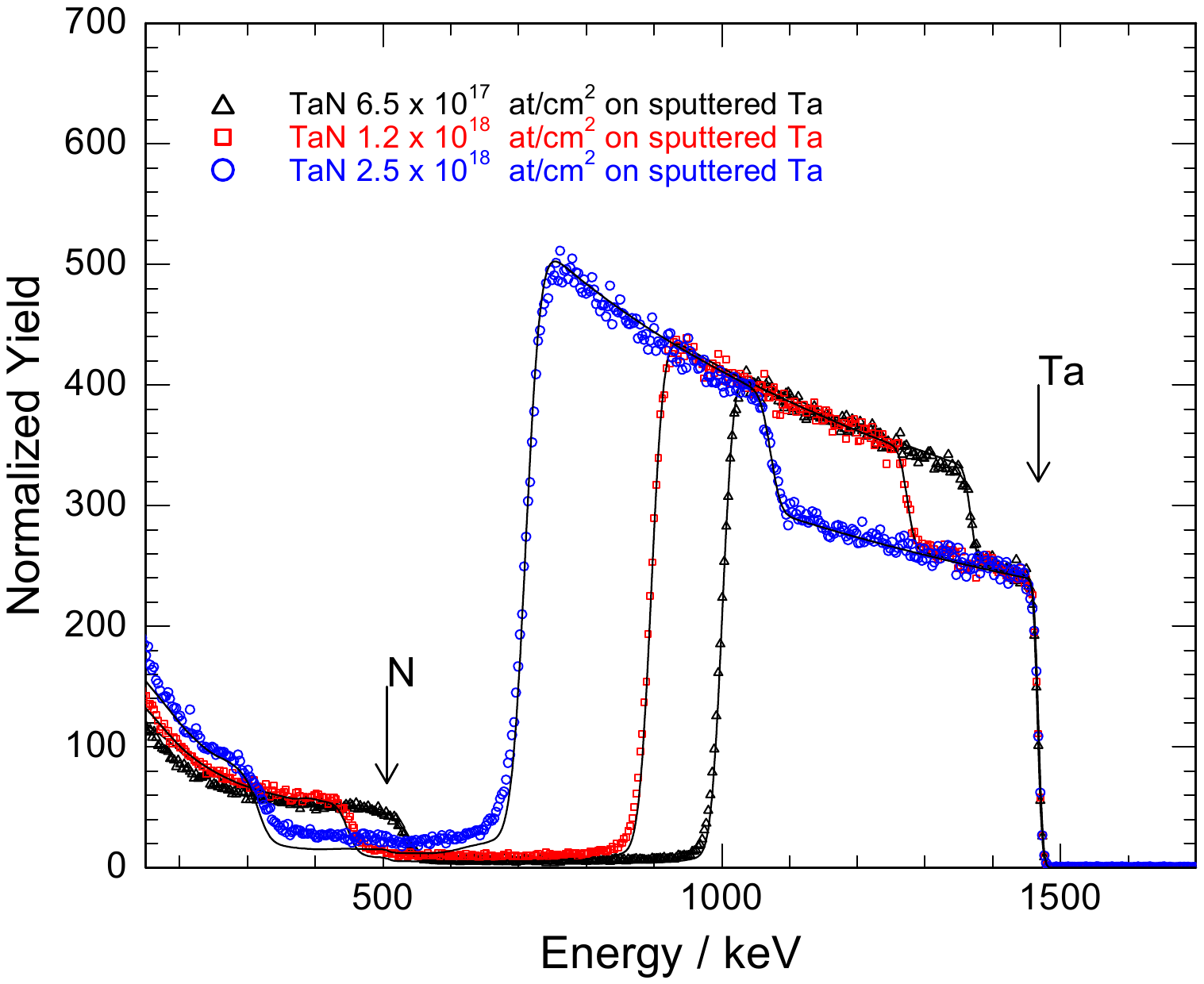}	
	\caption{RBS experimental spectra of the witness TaN/Ta samples deposited on silicon, measured at the Legnaro AN2000 accelerator (\num{1.59} MeV $^4$He$^+$ beam, $\theta_{\textup{out}}=\num{160}^\circ$). The SIMNRA 7 simulations superimposed to the experimental spectra (solid lines) refer to a stoichiometry ratio N/Ta=1 constant throughout the nitride layer.} 
	\label{fig:RBS_sputtered}%
\end{figure}

With the sputtering method, a series of targets with three different thickness were produced, ranging from \num{6.5e17} atoms/cm$^2$ to \num{2.5e18} atoms/cm$^2$ (Fig.~\ref{fig:RBS_sputtered}).
These correspond to energy losses between 15 and 40 keV, at a proton beam energy of 280 keV, as calculated using SRIM2013~\cite{ZIEGLER20101818}.
 
\begin{figure}[htb]
	\centering 
	\includegraphics[width=0.45\textwidth]{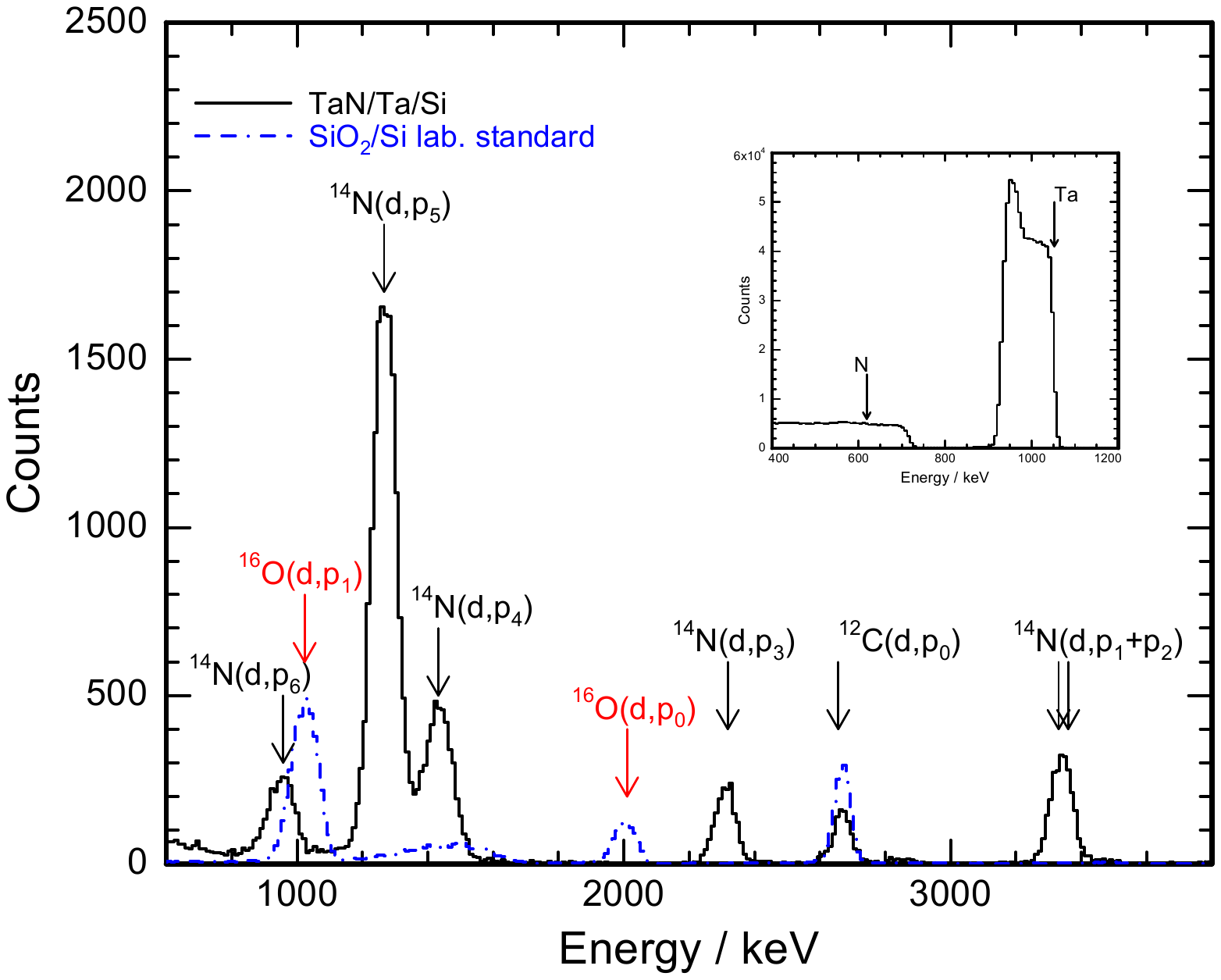}	
	\caption{NRA \num{1100}~keV (d,p) spectrum of a TaN/Si layer (black) superimposed to a SiO$_2$/Si laboratory standard (blue). The relevant $^{14}$N$(d,p_{i, i=1-6})$ peaks and the surface contamination $^{12}$C$(d,p_{0})$ peak are indicated. The $^{16}$O$(d,p_{0})$ peak in the TaN layer is depressed below the background of the measurement indicating that the concentration is below 1\%. The inset shows the RBS (d,d) spectrum of the tantalum nitride layer recorded simultaneously to the NRA spectrum at a scattering angle of 170$^\circ$.} 
	\label{fig:NRA}%
\end{figure}

The reactive sputtering process calibration, in terms of film stoichiometry and deposition rate, was established through multiple depositions on silicon wafers using natural nitrogen as the reactive gas, timely characterized by RBS and SEM (Scanning Electron Microscopy) analysis. 
To prove the beneficial effect of active gettering during deposition, the $^{16}$O$(d,p_{0,1})$ and $^{14}$N$(d,p_{i=1-6})$ nuclear reactions were used on a selected number of samples to determine the oxygen and nitrogen content in the films~\cite{wang2010handbook}. 
The analyses were performed at the CN accelerator of the Laboratori Nazionali di Legnaro. 
The NRA analyses (Fig.~\ref{fig:NRA}) showed that the oxygen content is below the minimum detectable limit of the technique (a fraction of \%) in the practical experimental conditions ($E_d=1100$~keV, $\theta = 150^\circ$, with \num{26} $\mu$m of aluminized Mylar in front of the \num{300}~mm$^2$ silicon detector). 
The $^{14}$N$(d,p_{i=1-6})$ reaction served to directly measure the nitrogen content in the films to confirm the RBS analyses.


The $\alpha$--RBS analysis was performed using a \num{1.59}~MeV $^4$He$^+$ beam at a backscattering angle $\theta_{\textup{out}}=\num{160}^\circ$ at the Legnaro AN2000 accelerator. The Si detector resolution was 12 keV (FWHM). The samples were measured with the beam impinging on the surface both at normal incidence and tilted at 5$^{\circ}$ with respect to the normal to the surface. The average ion beam current was kept below 1 pnA to guarantee a count rate $<$ 2 kHz to minimize pile-up effects. The beam spot was smaller than 1~mm$^2$. The RBS spectra were analyzed using  SIMNRA 7 \cite{10.1063/1.59188}.  The nitride layers were characterized by a constant stoichiometry throughout the coating and by a sharp interface towards the tantalum backing as expected for the production method. The stoichiometry resulted close to nominal 1:1 (±5\%) value of TaN. The deposition rate of tantalum nitride was \num{4.7e16} atoms/cm$^2$/min. The RBS spectra and the corresponding simulations are reported in Fig. \ref{fig:RBS_sputtered}.

\subsection{Target stability and beam-induced background}

\begin{figure}[htb]
	\centering 
	\includegraphics[width=0.5\textwidth]{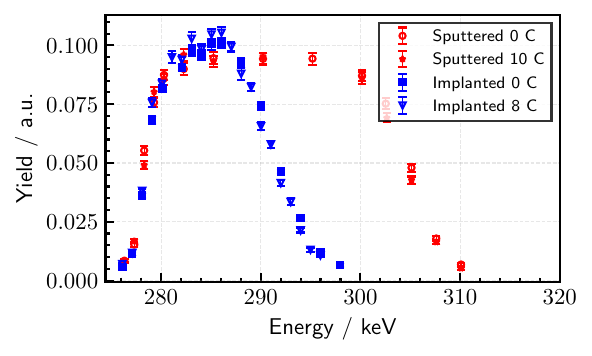}	
	\caption{Resonance scans measured at the LUNA-400 accelerator of the \num{278} keV resonance in \Npg\ for test targets produced with sputtering (blue) and implantation (red), before (squares) and after (stars) irradiation at the BIBF.} 
	\label{fig:scan_luna-400_2023}%
\end{figure}

Additional characterizations for the sputtered targets were carried out at the 2\,MV tandetron accelerator \cite {Rajta2018} of the ATOMKI accelerator center \cite{Biri2021} in Debrecen, Hungary. The thicknesses of the produced targets were checked with the resonance profile measurement of the \num{278}~keV resonance in \Npg. The beam induced background at various proton energies up to 1.7\,MeV
was also studied. Based on these experiments, some modifications of the target preparation technique were suggested, for example, the need for using enriched $^{14}$N gas in the sputtering process.

Target stability tests were performed on the produced samples at the LUNA-400~\cite{formicola_luna_2003} accelerator. Resonance scans of the targets were performed using an HPGe detector on the \num{278}~keV resonance. The samples were analyzed before and after irradiation with up to \num{10}~C collected charge at \num{360}~keV (Fig.~\ref{fig:scan_luna-400_2023}). No signs of degradation (within the 1-5\% statistical uncertainty of the measured points of the scans) was observed in these tests for the samples produced with the two different methods.

The amount of fluorine in the targets was also evaluated using the yield from the $^{19}$F(p,$\alpha\gamma$)$^{16}$O E$_p=$340~keV resonance. This element is a common source of BIB arising from contamination in the target backings and other materials exposed to the beam. 

\begin{figure}[htb]
	\centering 
	\includegraphics[width=0.5\textwidth]{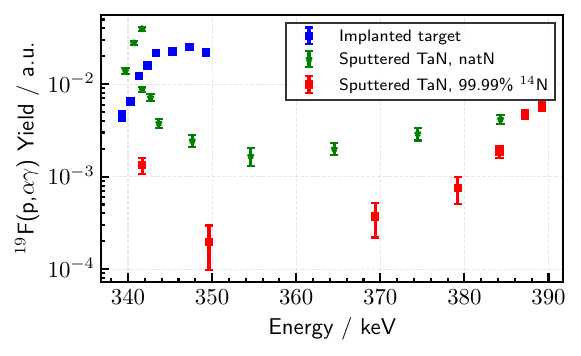}	
	\caption{Resonance scans of the $^{19}$F(p,$\alpha\gamma$)$^{16}$O E$_p=$340~keV resonance for different solid nitrogen target samples.} 
	\label{fig:19F_tests_yield}%
\end{figure}

In Fig.~\ref{fig:19F_tests_yield} sputtered natural nitrogen targets produced in an early test deposition are compared with a target produced with the \num{99.99}\% enriched $^{14}$N gas and with an implanted target. In the natural target (green) a fluorine contamination on the surface of the target was observed, as made evident by the higher yield at the $^{19}$F resonance energy, but a reduced amount is observed when the resonance is populated inside the sputtered material. This is in direct contrast with the yield observed in the implanted target (blue), here it is possible to observe a higher yield in the bulk of the target, coming from the fluorine contamination in the tantalum backing used to produce the target. The new sputtered samples (red) produced using enriched nitrogen gas showed much lower fluorine yield in the bulk of the sputtered material. For this target it was possible, given its thickness and the maximum available proton energy of $\sim$400~keV, to populate the resonance deep in the tantalum backing, where a significantly higher yield was observed with respect to the sputtered material.

In conclusion, these test performed at LUNA-400 underlined the quality and stability of the nitrogen targets and their readiness for the measurement, while the investigation of fluorine underlined its presence in Ta backing material.

\section{Experimental setup}
\label{sec:experimental}

\subsection{The Bellotti Ion Beam Facility}
\label{sec:BIBF}

This \Npg\ experiment was performed using the \num{3.5}~MV accelerator of the Bellotti IBF of Gran Sasso National Laboratories (LNGS)~\cite{junker_deep_2023}. Located in a deep-underground location, shielded by \num{1400}~m of rock, the experiments conducted in this environment can benefit from the reduction of the muonic component of the cosmic-rays background by six orders of magnitude. This directly translates into a reduction of the environmental component of the background observed in $\gamma$-ray spectra above \num{3.0}~MeV. 

The machine at the core of the facility is a single-ended electrostatic \num{3.5}~MV accelerator equipped with an ECR ion source and capable of providing an intense proton beam with up to 1~mA of current. This accelerator, developed with the stringent requirements needed for nuclear astrophysics investigations, features excellent terminal voltage stability (10~ppm) and a terminal voltage drift in the $10^{-5}$ range~\cite{sen_high_2019}.

During the measurements, a proton beam with $E_{p}=0.25-1.5$~MeV was delivered to the above described $^{14}$N solid targets, water-cooled and mounted at 90$^\circ$ with respect to the beam direction. A cold trap, made by a tantalum pipe mounted in front of the target was biased to -300~V and kept at liquid nitrogen temperature in order to suppress the secondary electrons and mitigate the carbon buildup on the target, respectively. The typical proton beam intensity was in the range of \num{200}-\num{400}~$\mu$A.

\subsection{Gamma detection setup}

\begin{figure}[htb]
	\centering 
	\includegraphics[width=0.45\textwidth]{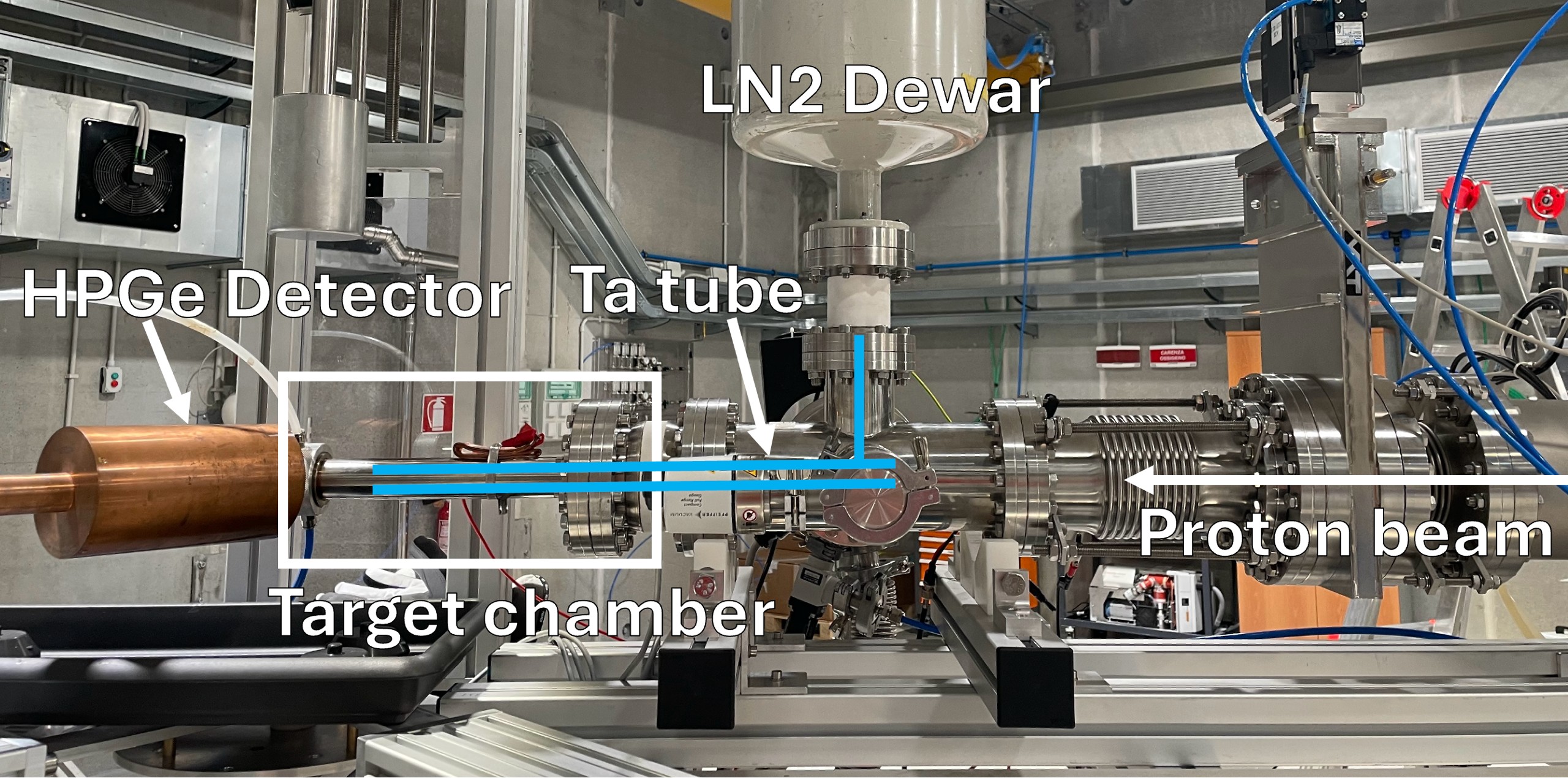}	
	\caption{A photo of the first, single detector experimental setup installed at the Bellotti IBF for the \Npg\ measurement. Some elements of the setup are labeled.} 
	\label{fig:setup}%
\end{figure}

During the initial phase of the \Npg\ cross section measurement, a High Purity Germanium (HPGe) detector with 120\% relative efficiency was installed. The detector was mounted at 55$^\circ$ with respect to the beam direction at a distance of \num{45}~mm from the target. In the subsequent data taking phase two other HPGe detectors were installed with nominal relative efficiency of 120\% and 60\%. In this configuration it was possible to move two HPGe detectors to cover four angles while the smaller 60\% detector was kept fixed vertically at 90$^\circ$ throughout the measurement. A photo of the first detection setup is shown in Fig.~\ref{fig:setup}.

A table was designed to hold the array of three HPGe detectors, each of them was fixed at predefined angles and moved on slides to change their distance from the target. This feature was also exploited to move the detectors farther away during target changes and to perform the efficiency calibration (see Sect.~\ref{sec:efficiency}).

During the angular distribution measurements, datasets were acquired at 0$^{\circ}$, 55$^{\circ}$, 90$^{\circ}$, 120$^{\circ}$ and 135$^{\circ}$. The detectors were kept at 10~cm away from their closest reference position from the chamber in order to minimize the systematic uncertainty of the angular distribution measurement.

\subsection{Detectors efficiency calibration}
\label{sec:efficiency}

\begin{figure}[htb]
	\centering 
	\includegraphics[width=0.52\textwidth]{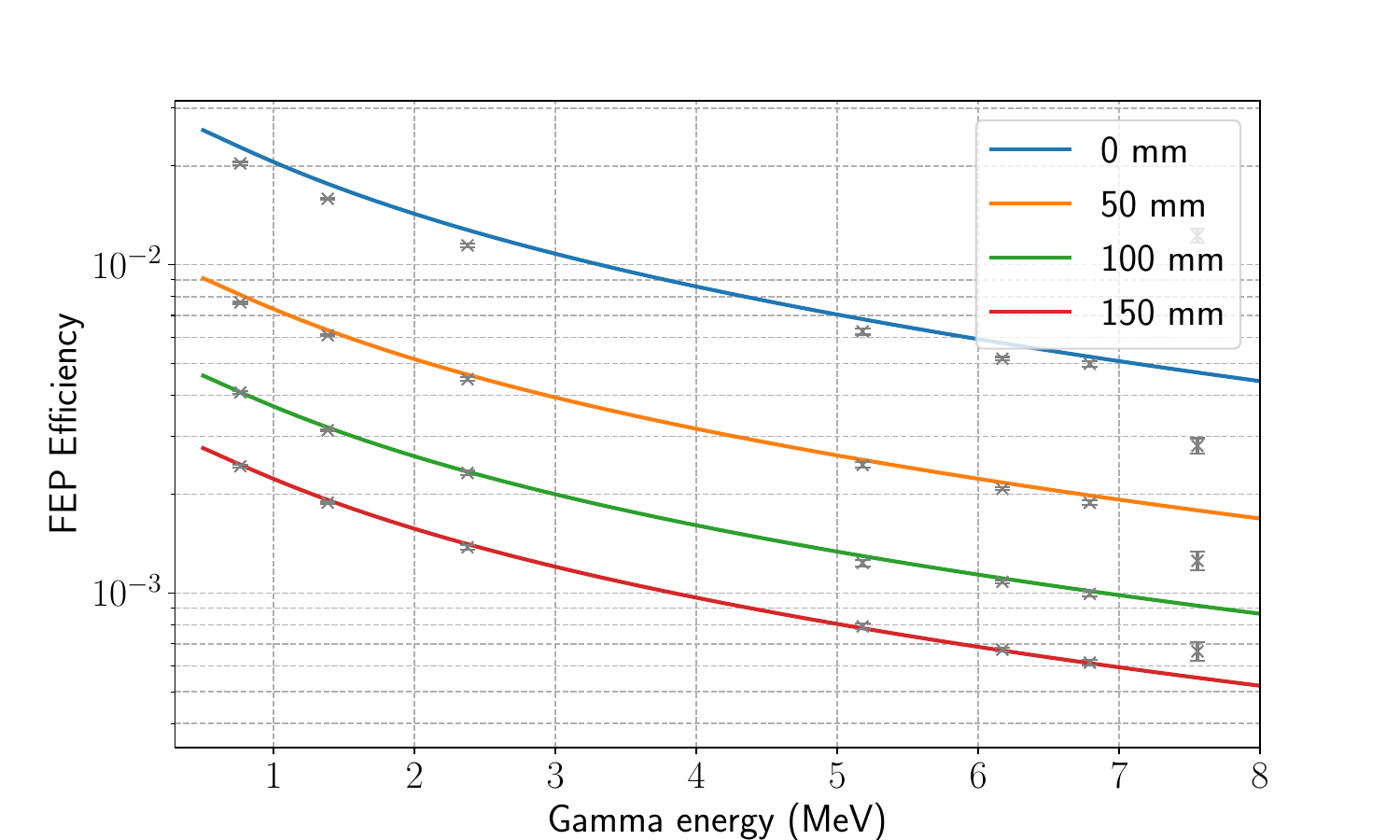}	
	\caption{Full-energy peak efficiency of one of the HPGe detectors at various distances. The points represent the measured values while the curves were obtained by a fit taking into account the summing effect, as described in the text. 
    } 
	\label{fig:efficiency}%
\end{figure}

The HPGe detectors were calibrated in efficiency using standard calibration sources $^{137}$Cs and $^{60}$Co to fix the low-energy region. For higher $\gamma$-energies the $^{14}$N(p,$\gamma$)$^{15}$O reaction at the 278~keV resonance was used, where the $\gamma$-ray branching ratios are precisely known \cite{Marta2011}. 

All efficiency measurements were performed at different distances in order to investigate the summing-in contribution to the ground state transition and the summing-out contributions for the transitions to various excited states. The yields $Y_i$ of the $\gamma$-rays emitted in the de-excitation of the populated $^{15}$O levels, where $i = \num{6.18}, \num{6.79}$ and \num{5.18}~MeV and the yield to the ground state, $Y_{\textup{g.s.}}$, can be expressed, introducing the summing-in and summing-out correction, as
\begin{equation}
\begin{gathered}
Y_i^{\textup{pri}} = R b_i \eta^{\textup{FEP}}(E_i^{\textup{pri}})(1-\eta^{\textup{TOT}}(E_i^{\textup{sec}})), \\
Y_i^{\textup{sec}} = R b_i \eta^{\textup{FEP}}(E_i^{\textup{sec}})(1-\eta^{\textup{TOT}}(E_i^{\textup{pri}})), \\
Y_{\textup{g.s.}} = Rb_{\textup{g.s.}} \eta^{\textup{FEP}}(E_{\textup{g.s.}}) + R \sum_i b_{\textup{i}} \eta^{\textup{FEP}}(E_{\textup{i}}^{\textup{sec}}) \eta^{\textup{FEP}}(E_{\textup{i}}^{\textup{pri}}),
\end{gathered}
\label{eq:summing_out}
\end{equation}
where $R$ is the total number of reaction per unit charge, $b_i$ the branching ratio of the given transition and $\eta^{\textup{FEP}}$ and $\eta^{\textup{TOT}}$ the full-energy peak and total efficiency of the detector, respectively. These two functions can be expressed using a standard parametrization and obtained by performing a fit of the experimentally observed yields measured at different distances from the target, as shown in \cite{imbriani_s-factor_2005}. 

The resulting efficiency curves as a function of $\gamma$-ray energy for one HPGe detector obtained with this approach are reported in Fig.~\ref{fig:efficiency}. The highest energy data points correspond to the ground state transition where the \linebreak summing-in effect is very strong, causing that these points lie above the curve. This indicates the necessity of the summing corrections discussed above.

\section{Selected results}
\label{sec:results}

\begin{figure}[tb]
\centering
  \includegraphics[width=0.5\textwidth]{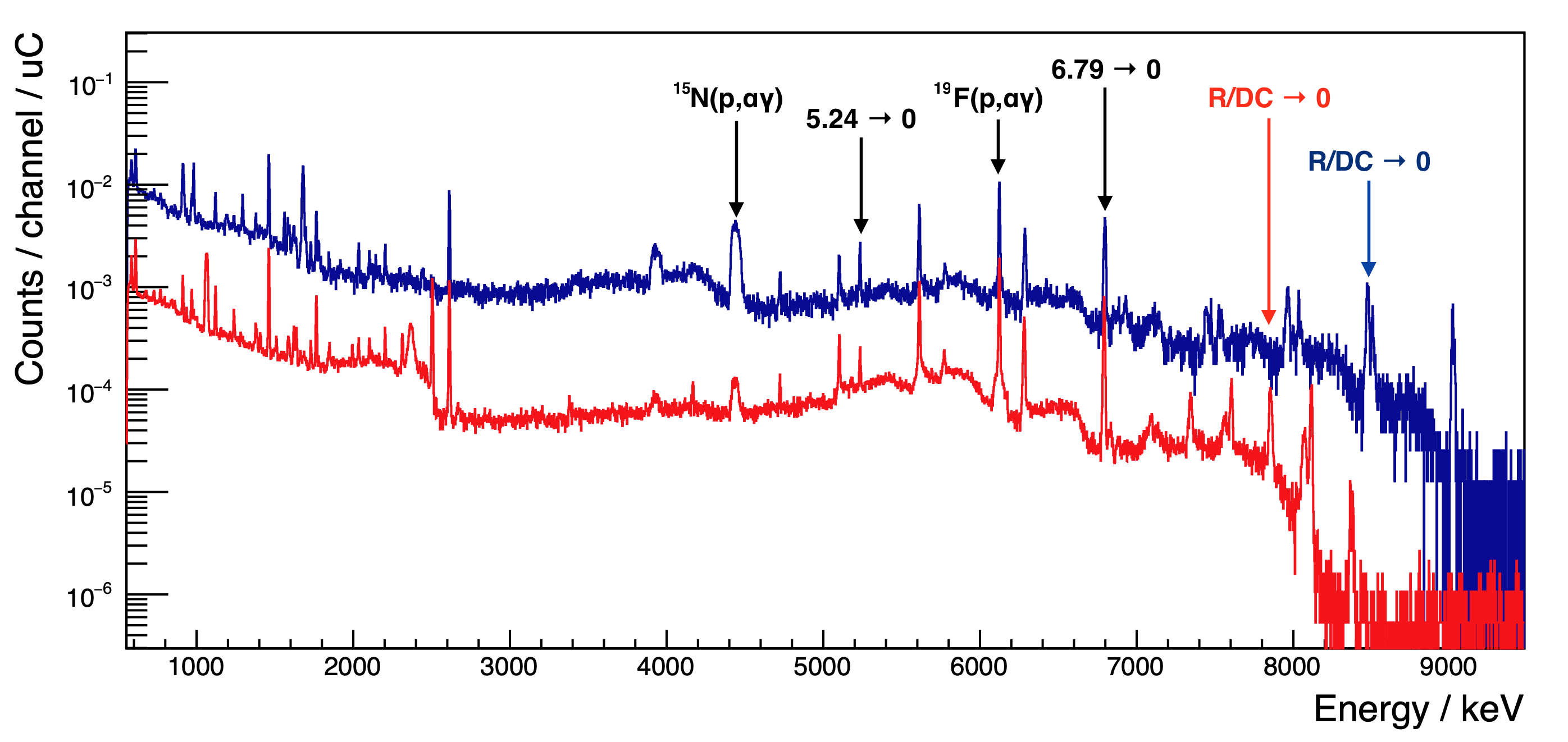}
\caption{Gamma-spectra measured in the first phase of the experiment at $E_p$ = 605~keV (red) and $E_p$ = 1261~keV (blue) on a sputtered target. The arrows show some peaks originating from the \Npg\ reaction and from beam-induced background.}
\label{fig:spectra}
\end{figure}

As described above, in the first phase of the measurements cross section data were collected with one HPGe detector placed at 55$^\circ$ in the energy range $E_{p}=0.25-1.5$~MeV. In the second phase, data at five angles (0$^\circ$, 55$^\circ$, 90$^\circ$, 120$^\circ$ and 135$^\circ$) were collected between $E_{p}=0.4-1.0$~MeV. Figure \ref{fig:spectra} shows two spectra from the first phase measured at $E_p$ = 605~keV (red) and $E_p$ = 1261~keV (blue) on a sputtered target. Some peaks originating from the \Npg\ reaction and from beam-induced background are labeled. 

\begin{figure}[tb]
\centering
  \includegraphics[width=0.45\textwidth]{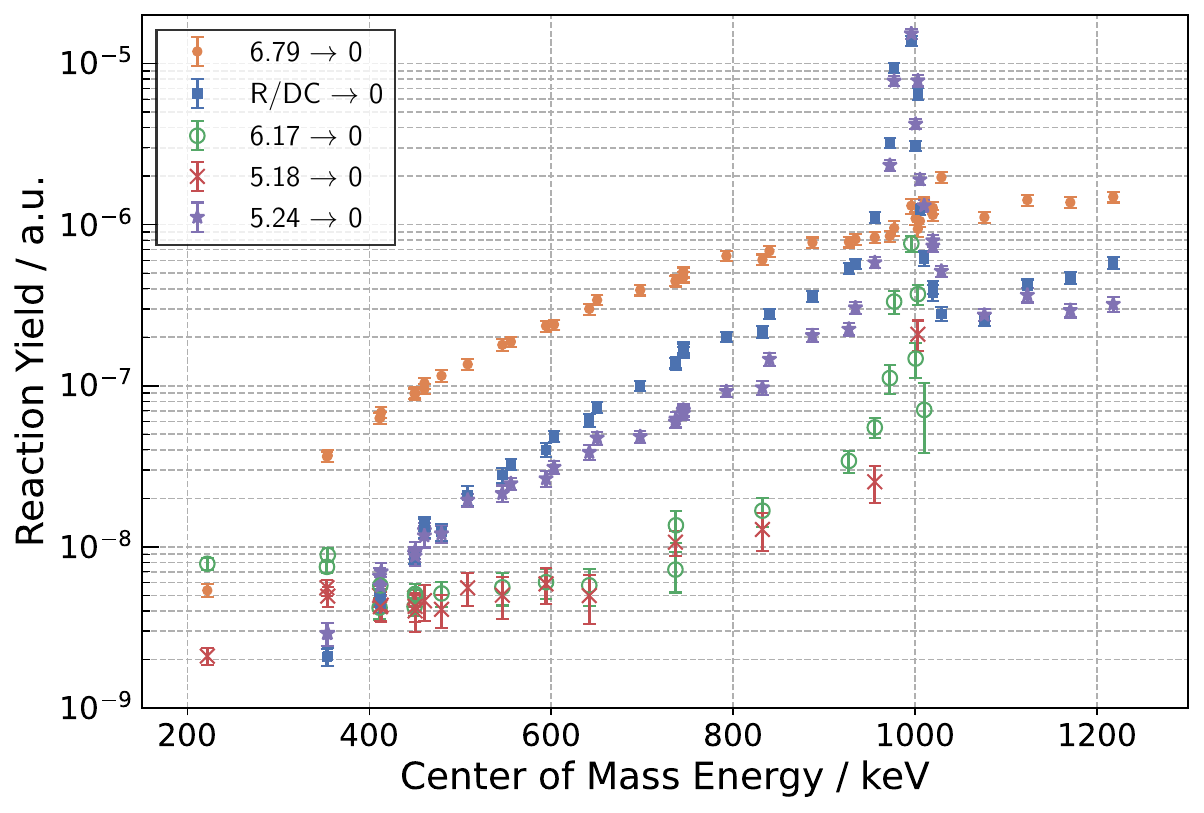}
\caption{Reaction yield corrected for target thickness and summing effects, for several $^{14}$N(p,$\gamma$)$^{15}$O transitions, obtained with the HPGe detector placed at 55$^\circ$.}
\label{fig:cross_section}
\end{figure}

Preliminary results from the first phase in the form of experimental yield (corrected for the target thickness and summing effects) are presented for the \num{6.79}~MeV, \num{6.17}~MeV, \num{5.18}~MeV, \num{5.24}~MeV secondary transitions and for the ground state in Fig.~\ref{fig:cross_section}. These results were obtained using sputtered targets. Notably, for the first time since the work of Schr\"oder~\emph{et~al.} \cite{SCHRODER1987240} it was possible to measure the weaker transitions in this energy range.

These preliminary results showcase the strength of the Bellotti IBF for studying the \Npg\ reaction, taking advantage of the high current and excellent long-term stability of the beam produced by the 3.5~MV accelerator, paired with the extremely low background conditions present in its underground location.

In the second phase, the angular distribution measurement focused on the most important transitions (R/DC$\rightarrow 6.79,0$), most notably moving down to 400~keV where no literature data are available. 
The results of these angular distribution experiment as well as the final data from the first phase will be presented in a forthcoming publication. 

\section{Conclusion and outlook}
\label{sec:conclusions}

In this paper, details of the experiment aiming at a new measurement of the \Npg\ reaction have been presented. Two different methods were used to produce high quality solid state nitrogen targets in order to reduce target-related systematic uncertainties. Several techniques were utilized to characterize these targets and to optimize the production procedure. Targets featuring low beam-induced background and high stability resulted from these efforts.

An experimental setup suitable for the Bellotti Ion Beam Facility has been designed and constructed. The setup is capable of holding three HPGe detectors which can be moved in order to measure angular distributions. The absolute efficiency of these detectors were measured including the detailed study of the summing effect, crucial especially for the highly significant ground state transition.

In first and second phases of the experiment, excitation functions at 55$^\circ$ detection angle for several transitions, and angular distributions for some selected transitions were measured, respectively. The analysis of these measurement is in progress. 

All of these new datasets will be included in a multi-channel R-matrix calculation where they are expected to play a significant role in constraining the extrapolation to astrophysical energies. Along with the experimental results, such an analysis will be included in a forthcoming publication. 

\section*{Acknowledgments}

This work was supported by HUN-REN Researcher Mobility Program 2023; by NKFIH grant K134197; by Fundação para a Ciência e Tecnologia (FCT, Portugal) through national funds to the Associated Laboratory in Translation and Innovation Towards Global Health REAL (LA/P/0117/2020); by the EU (ChETEC-INFRA, 101008324). E.M. acknowledges a fellowship by the Alexander von Humboldt Foundation. The work of the CAD service (M. Mongelli) and the mechanical workshop (C. Pastore, S. Martiradonna, N. Lacalamita and M. Franco) of INFN Bari is acknowledged.

%

%
%

\end{document}